\documentstyle[11pt,aaspp,flushrt]{article}
\newcommand{\Ref}{\hangindent=20pt \hangafter=1 \noindent}
\newcommand{\StartRef}{\hyphenpenalty=10000 \raggedright}

\newcommand{\beq}{\begin{equation}}
\newcommand{\eeq}{\end{equation}}
\newcommand{\NarrowMargins}{
  \setlength{\oddsidemargin}{+0.3in}
  \setlength{\evensidemargin}{-0.0in}
  \setlength{\textwidth}{6.2in}
  \setlength{\topmargin}{-0.75in}
  \setlength{\textheight}{9.25in}   }
\catcode`\@=11 

\def\lsim{\mathrel{\mathpalette\@versim<}}
\def\gsim{\mathrel{\mathpalette\@versim>}}

\def\be{{b^2 \over 8 \pi}}
\def\ve{{1 \over 2} \rho u^2}
\def\Ht{H_\theta}

\def\gt{\gamma_t}

\def\@versim#1#2{\vcenter{\offinterlineskip
        \ialign{$\m@th#1\hfil##\hfil$\crcr#2\crcr\sim\crcr } }}
\catcode`\@=12 
\NarrowMargins
\input{psfig}

\begin{document}
\title{On the Energetics of Advection-Dominated Accretion Flows}
\author{Eliot Quataert\footnote{equataert@cfa.harvard.edu} and Ramesh
Narayan\footnote{rnarayan@cfa.harvard.edu}} 
\affil{Harvard-Smithsonian Center for Astrophysics, 60 Garden
St., Cambridge, MA 02138}
\medskip
\setcounter{footnote}{0}

\begin{abstract}
Using mean field MHD, we discuss the energetics of optically thin, two
temperature, advection-dominated accretion flows (ADAFs).  If the
magnetic field is tangled and roughly isotropic, flux freezing is
insufficient to maintain the field in equipartition with the gas.  In
this case, we expect a fraction of the energy generated by shear in
the flow to be used to build up the magnetic field strength as the gas
flows in; the remaining energy heats the particles.  We argue that
strictly equipartition magnetic fields are incompatible with {\em a
priori} reasonable levels of particle heating; instead, the plasma
$\beta$ in ADAFs ($\equiv$ gas pressure divided by magnetic/turbulent
pressure) is likely to be $\gsim 5$; correspondingly, the viscosity
parameter $\alpha$ is likely to be $\lsim 0.2$.

\

\noindent {\em Subject headings:} accretion, accretion disks -- 
hydromagnetics -- turbulence
\end{abstract}

\section{Introduction}
Recent work (Narayan \& Yi 1994, 1995ab; Abramowicz et al. 1995; Chen
et al. 1995; Nakamura et al. 1997; Manmoto, Mineshige, \& Kusunose
1997; see Narayan, Mahadevan, \& Quataert 1998b, or Kato, Fukue, \&
Mineshige 1998 for reviews) has revived interest in a class of hot
optically thin accretion solutions first discovered by Ichimaru
(1977): advection--dominated accretion flows (ADAFs).  In ADAFs, the
accreting gas is unable to cool efficiently and most of the energy
generated by turbulent stresses is advected onto the central object.
As a result, the gas heats up to nearly virial temperatures and adopts
a two-temperature configuration (Shapiro, Lightman, \& Eardley 1976;
Rees et al. 1982), with the ions significantly hotter than the
radiating electrons.

In this paper, we derive and discuss the energy equations for
particles, magnetic fields, and turbulence in a two-temperature ADAF
(\S2).  There are several mutually exclusive prescriptions in the
literature for treating the energetics of ADAFs (compare, e.g., Esin,
McClintock, \& Narayan 1997, hereafter EMN and Manmoto et al. 1997).
The differences between these prescriptions lead to significant
differences in the predicted spectra of accreting black holes
(compare, e.g., Narayan et al. 1998a and Manmoto et al. 1997 for Sgr
A*), so it is important to resolve these discrepancies.  This is
undertaken in Appendix A.

Non-relativistic ADAFs exhibit a curious tension between two
conflicting requirements.  On the one hand, energy advection causes
the flow to be virial, with a profile of density and temperature ($T
\propto \rho^{2/3}$) that makes a non-relativistic gas nearly
isentropic.  On the other hand, dissipation of turbulent energy should
cause the gas entropy to increase inwards.  This tension is
particularly relevant for optically thin, two temperature, ADAFs which
have seen the most success in applications to real systems (see
Narayan et al. 1998b for a review); for such flows, $T_i \gg T_e$ and
the gas is quite accurately approximated as non-relativistic.

In an effort to alleviate this difficulty, Narayan \& Yi (1995b; NY95)
suggested that a proper treatment of the flow dynamics should
explicitly (if phenomenologically) account for the turbulence/magnetic
fields in the flow (since these components are not expected to behave
like non-relativistic particles). Using the energy equations derived
in \S2, we explore this suggestion in some detail.  In \S3 we show
that the most straightforward implementation of NY95's suggestion (a
constant ``$\beta$'' for the flow, which is utilized by many workers
in the field) is still incompatible with the turbulent heating of
particles. In \S4 we suggest an alternative method of including
turbulence in the dynamics of ADAFs which explicitly accounts for
particle heating; we present analytical and numerical solutions
consistent with this scenario.  Finally, in \S5 we summarize and
discuss the implications of this work.

\section{The Energetics of ADAFs}

The goal of this section is to derive energy equations for particles,
magnetic fields, and turbulence in a two-temperature ADAF (see \S 18
of Kato et al. (1998) for a related derivation).  Later in this paper
(and in Appendix A) we use these equations to clarify several issues
relevant to spectral and dynamical models of ADAFs.

Strictly speaking, since ADAFs are nearly collisionless, we should use
the Boltzmann equation to analyze their dynamics/energetics.  We
assume, however, that the use of MHD is justified by the presence of
the (turbulent) magnetic field.  In particular, this assumes that
various wave-particle interactions (pitch angle scattering) maintain
approximate isotropy in the distribution functions.  With this
simplification, the equations governing the structure of the accretion
flow are \beq {\partial \rho \over \partial t} + \nabla \cdot (\rho
{\bf v}) = 0,
\label{mass} 
\eeq \beq \rho { \partial {\bf v} \over \partial t} + \rho ({\bf v}
\cdot \nabla) {\bf v} = -\nabla P + \rho {\bf g} + {1 \over 4 \pi}
(\nabla \times {\bf B}) \times {\bf B} + \eta_v \nabla^2 {\bf
v},\label{mom} \eeq \beq { \partial {\bf B} \over \partial t} = \nabla
\times ({\bf v} \times {\bf B}) + \eta_B \nabla^2 {\bf B},\label{ind}
\eeq \beq \rho T{d s \over d t} = H - q^-, \label{energy} \eeq where
$\rho, {\bf v}, P,{\bf g}, {\bf B}, s, \eta_v,$ and $\eta_B$ represent
the mass density, velocity, gas pressure, gravitational acceleration,
magnetic field strength, particle entropy per unit mass, viscous
diffusivity, and magnetic diffusivity in the flow (respectively).  In
equations (\ref{mass})-(\ref{energy}), $d/dt$ denotes a Lagrangian
derivative while $\partial/\partial t$ denotes an Eulerian derivative.
Equation (\ref{energy}) is the sum of the electron and ion energy
equations, $H$ is a volumetric particle heating rate (due to viscous,
resistive, or plasma dissipation mechanisms) and $q^-$ is the
volumetric cooling rate.  Below we will discuss the individual ion and
electron energy equations.

We emphasize that $H$ is the heating rate of the particles.  In ADAFs,
$H$ is not necessarily equal to $q^+$, the rate at which energy is
generated by turbulent stresses/shear in the accretion flow, since
some of $q^+$ may be used to build up the magnetic field strength and
turbulent kinetic energy as the gas flows in (see \S2.3).  Similarly,
$s$ in equation (\ref{energy}) is not the total entropy of the fluid,
but only that of the particles.

We use mean field theory to derive energy equations for the turbulent
flow (see, e.g., Parker (1979) or Speziale (1991) for general
discussions of mean field theory and turbulent closure schemes; Balbus
\& Hawley (1998) and Kato et al. (1998) give nice applications of this
formalism to accretion theory).  We assume that each quantity
characterizing the flow can be expressed as a sum of an
ensemble-averaged (``mean'') quantity and a fluctuating quantity,
e.g., ${\bf v} = {\bf U} + {\bf u}$, ${\bf B} = {\bf B_0} + {\bf b}$,
$P = P_0 + p$, where the fluctuating quantities (${\bf u},{\bf b}, p$)
have zero mean (we denote ensemble-averages by $\langle \rangle $).
We assume that the turbulence is roughly incompressible, but we do not
take $\nabla \cdot {\bf v} = 0$.  Rather, we make the minimal
modification to the incompressibility assumption which allows us to
treat the compressive heating of particles/field/turbulence; we thus
take a static density field, $\rho = \rho({\bf r})$ (i.e., we take
$\partial_t \rho = 0$, but not $d_t \rho = 0$).  Another
simplification of this work is that we take ${\bf B_0} = 0$, i.e., we
assume that there is no mean magnetic field.

ADAFs are a one scale problem; the vertical scale height ($H_v$) is of
order the local radius in the flow and the inflow time of the gas is
of order the rotational period ($\Omega^{-1}$).  Assuming that
turbulent eddies have sizes $\sim H_v$ and turnover times $\sim
\Omega^{-1}$, this makes the use of a mean field theory somewhat
suspect.  For now, however, it is the best we can do.

By subtracting the mean versions of equations (\ref{mom}) and
(\ref{ind}) from the full equations, one derives equations for $u_i$
and $b_i$, the fluctuating velocity and magnetic field.  Multiplying
these equations by $u_j$ and $b_j$, respectively, yields equations for
the evolution of the Maxwell and Reynold's stress tensors.  The
resulting energy equations for the fluctuating magnetic field and
turbulent kinetic energy are given by (see also \S18 of Kato et
al. 1998)
\begin{eqnarray}
d_t \langle b^2\rangle &=& -2 \langle b^2\rangle \partial_k U_k + 2
{\langle b_i b_k\rangle } \partial_k U_i \nonumber \\ &-& \partial_k
\left(\langle u_k b^2\rangle - \eta_B \partial_k \langle b^2\rangle \right)
\nonumber \\ &+& 2 \langle b_i b_k \partial_k u_i\rangle -  \langle
b^2 \partial_k u_k\rangle - 2 \eta_B \langle ({\partial_k
b_i)^2\rangle } \label{ben}
\end{eqnarray}
and
\begin{eqnarray}
d_t \langle \rho u^2\rangle &=& -\langle \rho u^2\rangle \partial_k
U_k - 2 {\langle \rho u_i u_k\rangle } \partial_k U_i \nonumber \\ &-&
2 \partial_k \left({1 \over 2} \langle \rho u^2 u_k \rangle + \langle
p_{tot} u_k\rangle - \left \langle {b_i b_k u_i \over 4 \pi} \right
\rangle - {\eta_v \over 2} \partial_k \langle u^2\rangle \right)
\nonumber \\ &-& 2 \left \langle {b_i b_k \over 4 \pi} \partial_k u_i
\right \rangle + 2 \langle p_{tot} \partial_k u_k\rangle - 2 \eta_v
\langle ({\partial_k v_i)^2\rangle },
\label{ven}
\end{eqnarray}
where $p_{tot} = b^2/8\pi + p$ is the total fluctuating pressure and
$d_t = \partial_t + U_k \partial_k$ is the Lagrangian derivative with
respect to the mean flow.  The physical meaning of these terms is as
follows.  The first line on the right hand side of equations
(\ref{ben}) and (\ref{ven}) represents the generation of energy via
compression and coupling to shear in the mean flow.  The second line
gives the divergence of the flux of energy due to turbulence and
microscopic resistivity/viscosity.  The three terms on the third line
represent exchange of energy between the velocity and magnetic
components of the turbulence, fluctuating $PdV$ work due to
correlations (if present) between the fluctuating pressure and the
divergence of the turbulent velocity field, and dissipation by
microscopic resistivity/viscosity, respectively.

Taking the mean of equation (\ref{energy}) yields the mean energy
equation for the particles,
\begin{eqnarray}
d_t \langle \epsilon\rangle &=& - \gamma \langle \epsilon\rangle
\partial_k U_k + \langle H\rangle - \langle q^-\rangle \nonumber \\
&-& \partial_k\left({\langle p u_k\rangle \over \gamma - 1}\right) -
\langle p \partial_k u_k\rangle , \label{pen}
\end{eqnarray}
where $\epsilon = P/(\gamma - 1)$ is the thermal energy per unit
volume (and hence $\langle \epsilon\rangle = P_0/(\gamma - 1)$) and
$\gamma$ is the adiabatic index of the electron + ion gas; if both
species are non-relativistic, $\gamma = 5/3$, if both are
relativistic, $\gamma = 4/3$, etc..\footnote{In equation (\ref{pen})
we have written the particle entropy gradient as $\langle \rho T d_t s
\rangle = d_t \langle \epsilon \rangle - \gamma \langle \epsilon
\rangle U_k \partial_k \ln \rho$.  We use this interchangeably with
$\langle \rho T d_t s \rangle = \rho d_t \left \langle {c^2_s/(\gamma
- 1)} \right \rangle - \langle c^2_s \rangle U_k \partial_k \rho,$
where $c^2_s = P/\rho$ is the sound speed.}

A number of different ``$\gamma$'s'' will be used in this paper:
$\gamma$ refers to the adiabatic index of the electron + ion gas,
while $\gamma_e$ and $\gamma_i$ refer to the individual electron and
ion adiabatic indices, respectively.  The quantity $\gamma_t$ refers
to the ``adiabatic index'' of the turbulence, which is a measure of
how efficient flux freezing/compression is at increasing the turbulent
energy (\S4).  Finally, $\gamma_g$ refers to the total adiabatic index
of the fluid, containing contributions from the particles and
turbulence (\S2.2).

\subsection{The Total Energy Equation}

An equation describing the total energetics of the magnetic field,
turbulence, and particles is obtained by summing equations
(\ref{ben})-(\ref{pen}), which yields \beq d_t E + \partial_k F_k = G
- \langle q^-\rangle \label{tote} \eeq where \beq E = \left\langle \ve
\right\rangle + \left\langle \be \right\rangle + \langle
\epsilon\rangle \eeq is the total (turbulent plus thermal) energy per
unit volume,
\begin{eqnarray}
F_k &=& \left \langle {u_k b^2 \over 4 \pi} \right \rangle + {1 \over
2} \langle \rho u^2 u_k \rangle + {\gamma \over \gamma - 1} \langle p
u_k\rangle \nonumber \\ &-& \left \langle {b_i b_k u_i \over 4 \pi}
\right \rangle - {\eta_v \over 2} \partial_k \langle u^2\rangle -
{\eta_B \over 8 \pi} \partial_k \langle b^2 \rangle
\end{eqnarray} is a turbulent flux (the viscous/resistive contribution
is negligible), \beq G = -\partial_k U_k\left(\left \langle \ve \right
\rangle + 2 \left \langle \be \right \rangle + \gamma \langle
\epsilon\rangle\right) - T_{ik} \partial_k U_i \label{totg} \eeq is
the term associated with generating energy, and \beq T_{ik} = \langle
\rho u_i u_k\rangle - { \left \langle b_i b_k \right \rangle \over 4
\pi} \label{stress} \eeq is the stress tensor.

\subsection{Approximations}

Equations (\ref{ben})-(\ref{stress}) are fairly rigorous energy
equations for the particles/fluctuations.  They are, however, too
complicated for some of the calculations of interest to us (e.g.,
calculating the dynamics of an accretion flow).  We therefore make a
number of approximations to the mean field energy equations which make
them more suitable for simple analytical or numerical calculation.
These simplified equations maintain (hopefully) a number of the
essential features of the mean field equations.

First, we must provide a closure relation for the third order
turbulent quantities (i.e., the turbulent flux).  We choose, without
much justification, to ignore these terms, setting $F_k = 0$.  Second,
we assume that the diagonal components of the turbulent stress tensor
are equal, in the sense that $\langle b^2_r\rangle \approx \langle
b^2_\theta\rangle \approx \langle b^2_\phi\rangle $, with analogous
equations for $\langle v^2_i\rangle $.  We do not, however, ignore the
off diagonal terms in the stress tensor since they are responsible for
angular momentum transport.  The generalization to arbitrary diagonal
components for the stress tensor is given in Appendix B.

We also assume that the flow is azimuthally symmetric, has $U_\theta =
0$, and rotates uniformly on spherical shells (i.e., $U_\phi = r
\Omega(r) \sin \theta$, where the rotation rate $\Omega$ is
independent of $\theta$; see Narayan \& Yi 1995a).  Taking $U_\theta =
0$ requires that the angular scale height of the flow be roughly
independent of radius (Narayan \& Yi 1995a; Abramowicz et al. 1997).
With these approximations, we can write, in spherical coordinates,
\beq G = - \left( {5 \over 3} \left \langle \ve \right \rangle + {4
\over 3} \left \langle \be \right \rangle + \gamma \langle
\epsilon\rangle \right) \partial_k U_k - T_{r \phi} \sin \theta r {d
\Omega \over d r} - T_{\theta \phi} \Omega \cos \theta.
\label{gsimple} \eeq This equation demonstrates that, under compression,
the velocity component of the turbulence behaves like a gas of
adiabatic index $5/3$ while the magnetic field behaves like a gas of
adiabatic index $4/3$.  This is a consequence of our assumption of
isotropic turbulence.

We express the level of turbulence in the flow in terms of the gas
pressure via two parameters: \beq \beta_b = {\langle P\rangle \over
\langle b^2/8 \pi\rangle} \ \ \ \ {\rm and} \ \ \ \ \beta_v = {\langle
P\rangle \over \langle \ve\rangle}. \label{beta} \eeq We generally
take $\beta_b = \beta_v \equiv \beta$ (i.e., the turbulence is
``Alfvenic'').  Our definition of $\beta$ is that utilized in the
plasma physics literature.  A number of workers in the accretion
literature (e.g., NY95) define a ``$\beta$'' via $\beta_{\rm adv}
\equiv P_{gas}/P_{tot}$, i.e., the fraction of the total pressure
supplied by the gas.  This is related to our $\beta$ by $\beta_{\rm
adv} = 3\beta/(3\beta + 1)$ or $\beta_{\rm adv} = \beta/(\beta + 1)$,
depending on whether one defines the magnetic pressure to be $b^2/24
\pi$ or $b^2/8 \pi$.

Finally, we set $\theta = \pi/2$, i.e., midplane values, in the above
expressions; this is a crude -- but standard -- form of height
integration.  With these assumptions it is rather straightforward to
show that the total energy equation for the flow (eq. [\ref{tote}])
takes the form (dropping the $\langle \rangle $ for simplicity) \beq
\rho U_r \partial_r \left({ v^2_t \over \gamma_{g} - 1}\right) = U_r
v^2_t \partial_r \rho + q^+ - q^-,
\label{esimple} \eeq where $v_t$, the effective isothermal sound
speed, is related to the real isothermal sound speed of the gas
($c_s$) and the turbulent sound speed ($c_t$) by \beq v^2_t = c^2_s +
c^2_t = c^2_s(1 + \beta^{-1}), \ \ \ c^2_s = {\langle P \rangle \over
\rho}, \ \ \ c^2_t = {\langle b^2 \rangle \over 8 \pi \rho} = {c^2_s
\over \beta}.\label{speed} \eeq The quantity $\gamma_g$ is an
effective adiabatic index for the flow which includes contributions
from the particles, the magnetic field, and the turbulence.  For
isotropic Alfvenic turbulence, we have \beq \gamma_g = {\gamma + 3
(\gamma - 1) \beta^{-1} \over 1 + 2 (\gamma - 1) \beta^{-1}}.
\label{gamma} \eeq  If we take the ion + electron gas to be
 nonrelativistic (a good approximation), equation (\ref{gamma})
simplifies to \beq \gamma_g = {5 \beta + 6\over 3 \beta +
4}.\label{gsimp}\eeq This expression for $\gamma_g$ is slightly
different from that of Esin (1997) who found, in our notation,
$\gamma_g = (15 \beta + 8)/(9 \beta + 6)$. This is because she
neglected the ram pressure of the turbulence, i.e., she set $\beta_v
\rightarrow \infty$.

The particular expressions for $v_t$ and $\gamma_g$ in equations
(\ref{speed}) and (\ref{gamma}) are a consequence of our assumption of
isotropic Alfvenic turbulence, the validity of which is highly
uncertain.  Appendix B generalizes these expressions to the case of
arbitrary diagonal components for the stress tensor.

The quantity $q^+$ in equation (\ref{esimple}) is the rate of energy
generation in the flow due to coupling between the turbulence and the
shear.  It is given by \beq q^+ = - T_{r \phi} r {d \Omega \over dr},
\eeq which can be ``closed'' using a Shakura \& Sunyaev (1973)-like
relation, \beq T_{r \phi} = - \alpha \rho v_t H_v r {d \Omega \over
dr} .\eeq

Equation (\ref{esimple}) demonstrates an important point.  In order to
write an energy equation with $q^+$ (the standard viscous heating
term) as a source term, one must consider all energy components in the
flow (hence the appearance of the ``effective'' sound speed and
``effective'' adiabatic index in eq. [\ref{esimple}], which contain
contributions from the turbulence and the particles).  It is, strictly
speaking, incorrect to write a particle energy equation with $q^+$ as
a source term, since the particle heating rate, $H$, is, in general,
somewhat different from $q^+$.  The relationship between the
``particle'' and ``total'' energy equations is discussed further in
Appendix A.

\subsection{The Electron, Ion, and Turbulence Energy Equations}

The mean electron and ion energy equations are given by the individual
versions of equation (\ref{pen}) which, utilizing the assumptions
given above (steady state, azimuthal symmetry, dropping third order
turbulent quantities, etc.), simplify to

\beq U_r \partial_r \epsilon_i = \gamma_i \epsilon_i U_r \partial_r
\ln \rho + H_i - q_{ie} - q^-_i \label{pe} \eeq and \beq U_r
\partial_r \epsilon_e = \gamma_e \epsilon_e U_r \partial_r \ln \rho +
H_e + q_{ie} - q^-_e, \label{ee} \eeq where $H_e$ ($H_i$) is the
energy dissipated out of the turbulence which heats the electrons
(ions), $\gamma_e$ ($\gamma_i$) is the monatomic ideal gas adiabatic
index for the electrons (ions), $q^-_e$ ($q^-_i$) is the electron
(ion) cooling rate, and $q_{ie}$ is the ion-electron energy exchange
rate.  We have dropped the $\langle \rangle$ for conciseness.

As it stands, however, equations (\ref{pe}) and (\ref{ee}) are not
closed since there is no prescription for calculating $H_e$ and $H_i$.
Let us define $\delta_H$ to be the fraction of the dissipated
turbulent energy which heats the electrons, so that $H_e = \delta_H H$
and $H_i = (1 - \delta_H)H$.  To get a handle on the particle heating
rate, $H$, we note that it is equal to the total energy dissipated out
of the turbulence (magnetic and velocity), and thus appears in the
energy equation for the turbulence.  Summing equations (\ref{ben}) and
(\ref{ven}), and employing the various simplifying assumptions
outlined in \S 2.2, we find that the energy equation for the
turbulence simplifies to \beq U_r \partial_r \left( \be + \ve\right) =
\left({4 \over 3} \be + {5 \over 3} {\rho u^2 \over 2}\right) U_r
\partial_r \ln \rho + q^+ - H,
\label{turbe} \eeq which becomes 
\beq 2 \rho U_r \partial_r c^2_t = c^2_t U_r \partial_r \rho + q^+ - H
\label{turbesimp} \eeq for isotropic Alfvenic turbulence (for which
$b^2/8\pi = \rho u^2/2$, $c^2_t = b^2/8\pi\rho$, and the turbulent
``adiabatic index'' is 1.5).

Equations (\ref{turbe}) and (\ref{turbesimp}) state that the rate of
change of the turbulent energy as the gas accretes inward is that
supplied by compression/flux freezing (the $\partial_r \rho$ term) and
coupling to the shear in the mean flow ($q^+$) minus the energy
dissipated into particle heat ($H$).  We can rewrite these relations
schematically as \beq Q_t = q^+ - H,\eeq where $Q_t$ is the rate of
change of the ``turbulent entropy.''  Note that $H$, the particle
heating rate, shows up as a cooling term in the turbulent energy
equation.

In ADAFs, just as one cannot ignore the advection term in the particle
energy equation, one cannot set $H = q^+$ and $Q_t = 0$ in the
turbulent equation.  We do, however, expect that $H \sim q^+$ (rather
than, say, $H \ll q^+$ or $H \equiv q^+$).  This is because of the one
scale nature of the turbulent energy equation: the timescale on which
shear generates energy (the rotational period) is also the timescale
on which energy cascades to small scales to heat particles (i.e., the
eddy turnover time), which is also $\sim$ the timescale on which
equipartition attempts to be established (the inflow time of the
gas).\footnote{By contrast, in thin accretion disks the inflow time of
the gas is much longer than the the rotational period. To an excellent
approximation, then, one can set $Q_t = 0$ and $H = q^+$.  This is the
turbulent analog of being able to drop the entropy gradient in the
particle energy equation, setting $q^- = H$ (which becomes the more
standard $q^- = q^+$ by the above argument).}

For the particular case of isotropic turbulence, $Q_t \gsim 0$ and
thus $H \lsim q^+$.  This is because, under compression, isotropic
Alfvenic turbulence behaves like a $\gamma = 1.5$ gas, i.e., $b^2
\propto \rho^{1.5}$ (cf eq. [\ref{turbesimp}]).  By contrast, for a
quasi-spherical flow, equipartition with the gas requires $b^2 \sim
\rho c^2_s \propto \rho^{5/3}$ (\S3).  Compression (flux freezing)
therefore does not supply sufficient energy to keep the turbulence in
equipartition with the gas.  Some of the energy generated by the
coupling between the turbulence and the mean flow ($q^+$) goes into
increasing the turbulent energy, while the remainder heats the
particles (hence $H \lsim q^+$).

For non-isotropic turbulence configurations, one could have $H \gsim
q^+$.  This occurs when the turbulence behaves like a gas of adiabatic
index $\gsim 5/3$ (see Appendix B).  Flux freezing/compression then
acts to increase the level of turbulence above the equipartition
value.  In this case, the turbulent energy dissipated in the flow
($H$) must exceed that generated by shear ($q^+$), in order for the
flow to remain gas pressure dominated (Bisnovatyi-Kogan \& Lovelace
1997).

Many models of two-temperature ADAFs in the literature parameterize
the turbulence by assuming a constant $\beta$ for the flow (note that
in deriving the above equations we have not employed this assumption).
In this case, one can substitute $c^2_t = c^2_s \beta^{-1}$ into
equation (\ref{turbesimp}) and, knowing the dynamics of the flow
(which gives $c^2_s$, $\rho$, $U_r$, $q^+$, etc., as a function of
$r$), calculate the particle heating rate that is implicitly assumed
by these models (since $H$ is the remaining unknown in equations
[\ref{turbe}] and [\ref{turbesimp}]).  This is carried out in the next
section; surprisingly, and problematically, it turns out that many
models in the literature require $H \ll q^+$, i.e., most of the energy
generated by turbulent stresses ($q^+$) is implicitly assumed to be
advected radially by the turbulence ($Q_t$), rather than heating the
particles ($H$).

\section{The Dynamics of ADAFs}
In this section we reanalyze current treatments for phenomenologically
including turbulence/magnetic fields in the dynamics of ADAFs.  We
show that explicitly decomposing the energetics of the flow into a
turbulent/magnetic component and a gas component, as done in the
previous section, leads to a significant reinterpretation of these
models.

Consider a steady axisymmetric accretion flow.  The dynamics of
viscous hydrodynamic accretion can (following Narayan, Kato, \& Honma
1997, hereafter NKH) be described by the following height integrated
equations representing conservation of mass, radial momentum, angular
momentum, and energy,

\beq {d \over dr}(\rho r^2 H_\theta U_r)= 0,
\label{masscons} \eeq \beq U_r {d U_r \over dr} - \Omega^2 r = -
\Omega_K^2 r - {1 \over \rho} \, {d \over dr } ( \rho v_t^2),
\label{radialeq} \eeq \beq U_r \, {d(\Omega r^2) \over dr} = {1 \over
\rho r^2 \Ht} \, {d \over dr} \left( {\alpha \rho v^2_t r^4 \Ht \over
\Omega_K} {d\Omega \over dr} \right), \label{angulareq} \eeq \beq \rho
U_r T {ds\over dr} = q^+ - q^-.
\label{energyeq}
\eeq Anticipating the quasi-sphericity of the flows of interest, we
have carried out height integration on spherical shells, so that
$H_\theta$ is the angular scale height of the flow; where necessary,
we take $\Ht = v_t/(r \Omega_K)$.\footnote{In spherical coordinates,
balancing pressure and the centrifugal force in the $\theta$ direction
implies $\rho \propto \exp\left[-0.5 \Omega^2 r^2
\cos^2\theta/v^2_t\right]$ (e.g., Narayan, Barret, \& McClintock
1997), assuming $v_t$ and $\Omega$ to be independent of $\theta$
(this, but {\em not} $\Omega$, $v_t$ independent of $z$, is a
reasonable approximation for ADAFs).  Defining $\Ht$ to be the
spherical average of the density, i.e., $\Ht =(4 \pi)^{-1} \int
\sin\theta d\theta d\phi \rho$ yields $\Ht = (\pi/2)^{1/2} x \ {\rm
erf}(x^{-1}/\sqrt2)$, where $x = v_t/(\Omega r)$.  This is the correct
expression for $\Ht$ for ADAFs.  It is, however, numerically
cumbersome, so for now we stick with the (more standard) expression
given in the text.}  

In the spirit of the previous section, we interpret the pressure
(sound speed) and entropy in equations
(\ref{masscons})-(\ref{energyeq}) as effective quantities, containing
contributions from both the particles and the turbulence.  In
particular, we take $v^2_t = c^2_s + c^2_t$ as the sound speed
(eq. [\ref{speed}]) and interpret equation (\ref{energyeq}) as
equation (\ref{esimple}) (with the effective adiabatic index
$\gamma_g$).  It is straightforward to employ the
averaging/approximation scheme discussed in the previous section to
the MHD momentum equation (eq. [\ref{mom}]) and derive equations
(\ref{radialeq}) and (\ref{angulareq}).

\subsection{Self-Similarity}

In the limit of $q^- \ll q^+$, cooling provides a rather minor
perturbation when calculating the dynamics of the flow.  One can
therefore set $q^- = (1-f)q^+$ in the energy equation (with $f \sim
1$) so that $f q^+$ of the viscous energy is advected by the flow.  In
this limit, NY94 showed that, if $\gamma_g$ and $f$ are independent of
radius, and $\Omega_K$ is Newtonian, equations
(\ref{masscons})-(\ref{energyeq}) admit a self-similar solution, with
the various flow variables being power laws in radius (note that this
entails that $\beta$ is a constant, independent of radius).  In
particular, $\Ht =$ constant, $U_r \propto \alpha r^{-1/2}$, $v^2_t
\propto r^{-1}$, $\rho \propto \alpha^{-1} r^{-3/2}$, and $\Omega
\propto (5/3 - \gamma_g)^{1/2} \ \Omega_K$ (see NY94 or Narayan et
al. 1998b for the complete scalings).

Of particular interest for our purposes is the dependence of $\Omega$
on $\gamma_g$.  Non-relativistic ($\gamma_g = 5/3$)
advection-dominated accretion flows tend to be non-rotating.  The
reason is that the above equations describing the structure of the
accretion flow in 1D exhibit the following ``singularity.''
Approximate hydrostatic equilibrium in the radial direction implies a
virial flow, for which $v^2_t \sim r^2\Omega_K^2 \propto r^{-1}$.
Together with angular momentum conservation, this implies $U_r \sim
\alpha r^{-1/2}$.  Mass conservation then implies $\rho \propto
\alpha^{-1} r^{-3/2}$.  Together, then, mass, radial momentum, and
angular momentum conservation impose a $T \propto v^2_t \propto
\rho^{2/3}$ density/temperature profile on the flow.  The remaining
flow variable, $\Omega$, adjusts itself to ensure that the energy
equation is satisfied.  The profile $T \propto \rho^{2/3}$ is,
however, an adiabat for a non-relativistic flow with $\gamma_g = 5/3$.
This means that there can be no entropy generation as the gas flows
inwards, which requires that $\Omega$ and $q^+$ go to zero.

The remainder of this paper is predicated on the assumption that this
singularity as $\gamma_g \rightarrow 5/3$ is, at some level, real, and
is not an artifact of the approximate treatment of the dynamics given
above.  Height integration schemes differing from ours do {\em not}
show this behavior (e.g., Chen et al. 1997; Manmoto et al. 1997).  We
believe that these models, based on cylindrical averaging, are not
accurate representations of the quasi-spherical flows of interest, but
only 2D or 3D calculations will fully resolve this issue.

NY95 proposed that non-relativistic ADAFs (such as the much applied
$T_i \gg T_e$ variety) behave differently from a $5/3$ gas because of
turbulence (in particular, magnetic fields) in the flow.  The
``formal'' justification for this suggestion is given by the
derivation of \S2 (cf eqs. [\ref{esimple}]-[\ref{gsimp}] and the
accompanying discussion).  In what follows, we employ the most
straightforward implementation of NY95's suggestion; we analyze the
dynamics of ADAFs assuming non-relativistic particles and a constant
$\beta$ turbulent/magnetic contribution to the pressure and energy
density of the flow.

In this case, the effective adiabatic index $\gamma_g \ne 5/3$, and
the flow can be differentially rotating.  This is important since we
expect real flows to possess non-negligible angular momentum.  If,
however, we take the self-similar solution for $\gamma_g \ne 5/3$ and
compute the entropy gradient for a nonrelativistic gas ($\gamma =
5/3$), we find $T d s/dr|_{\rm gas} = 0$.  Accounting for
turbulence/magnetic fields to make the flow as a whole ($\gamma_g$)
behave differently from a $5/3$ gas does not alter the fact that the
gas component of the flow still has $\gamma \approx 5/3$ (see Appendix
A), and therefore has no entropy gradient in the self-similar regime.
It therefore does not alleviate the fundamental tension between being
virial and the need for particle heating.

Using the energy equations given in \S2 (see the discussion below
eq. [\ref{turbesimp}]), it is straightforward to show that $H = q^- =
(1-f)q^+$ and $Q_t = fq^+$.  This implies that, although including
constant $\beta$ turbulence in the dynamics of a non-relativistic ADAF
allows for differential rotation and entropy generation ($q^+ \ne 0$),
it does so by requiring that almost none of the energy generated by
turbulent stresses heats the particles (since for $f \sim 1$, $H \ll
q^+$).  Instead, this energy is assumed to be stored as turbulent
energy in the flow, i.e., it is advected by the turbulence, rather
than the particles.  For the self-similar solution, this conclusion is
{\em independent} of $\beta$.  For larger $\beta$, the turbulent
energy is smaller; $\gamma_g$ is, however, closer to $5/3$ and so
$\Omega$ and $q^+$ are smaller as well.  The solution always adjusts
itself so that the energy generated by the shear is precisely what is
required to maintain constant $\beta$ in the flow.

The condition $H \ll q^+$ means that only a small fraction of the
turbulent energy cascades to small scales and heats the particles.
This seems unreasonable since intuition and dimensional analysis
dictate that ADAFs should have $H \sim q^+$ (although not $H \equiv
q^+$; see \S2.3).

\section{An Alternative Dynamics: Explicit Inclusion of Particle Heating}

One way to resolve the above contradiction is to give up the
assumption of constant $\beta$.  Instead, let us assume that the
magnitude of the turbulence/magnetic fields in the flow adjusts so as
to allow the requisite level of particle heating.

To analyze this scenario, we replace the single energy equation
representing ``particles plus turbulence'' (eq. [\ref{esimple}]) with
the following two energy equations, the first for the particles and
the second for the turbulence, \beq \rho U_r \partial_r \left({ c^2_s
\over \gamma - 1}\right) = U_r c^2_s \partial_r \rho + H - q^-
\label{partesimp}\eeq \beq \rho U_r \partial_r \left({ c^2_t \over
\gamma_{t} - 1}\right) = U_r c^2_t \partial_r \rho + q^+ - H.
\label{tesimp} \eeq The quantity $\gamma$ ($\gamma_t$) is the particle
(turbulent) adiabatic index.  For $c^2_t = b^2/8\pi\rho$ and $\gamma_t
= 1.5$, equation (\ref{tesimp}) was derived in \S2.3 (it is equivalent
to eq. [\ref{turbesimp}]); here we allow for a slightly more general
fluid model of the turbulence by considering a general value of
$\gamma_t$.  Equation (\ref{partesimp}) is the sum of the individual
particle energy equations (eqs. [\ref{pe}] and [\ref{ee}]).

To ensure particle heating, we replace the assumption of a constant
$\beta$ with the assumption that a constant fraction $\eta$ of the
energy generated by turbulent stresses heats the particles, i.e., \beq
\eta \equiv H/q^+ = {\rm constant}. \label{eta2} \eeq $\beta(r)$ is
now an output of the dynamical equations.  Mass, radial momentum, and
angular momentum conservation are still given by equations
(\ref{masscons})-(\ref{angulareq}), taking the total pressure to be
the sum of the gas and turbulent pressures, $v^2_t = c^2_s + c^2_t$.
To be consistent with previous treatments, we define $q^- = (1-f)q^+$
(although $q^- = (1-f)H$ might be a more physical paramatrization).
With these definitions, \beq H - q^- = (\eta + f - 1)q^+ \ \ \ {\rm
and} \ \ \ q^+ - H = (1 - \eta)q^+. \eeq

As discussed in Appendix B, a crucial assumption of the present
analysis is that $\gamma_t \lsim 5/3$.  This is equivalent to the
statement that the effect of compression/flux freezing on the
turbulent energy is insufficient to maintain the turbulence in
equipartition with the gas.  In this case, it is $q^+$, the coupling
between the turbulence and the shear, which is primarily responsible
for increasing the turbulent/magnetic energy as the gas accretes
inwards.

\subsection{Self-Similarity}

We search for analytical solutions to equations
(\ref{masscons})-(\ref{angulareq}) and
(\ref{partesimp})-(\ref{tesimp}), assuming that $f$, $\eta$, and
$\alpha$ are independent of $r$.  For simplicity we take $\alpha^2 \ll
1$ (almost certainly a reasonable approximation), $\beta \gsim 1$
(also likely to be a reasonable approximation), and $\gamma = 5/3$
(non-relativistic gas, i.e., ion dominated).  The assumption on
$\beta$ is needed in order to express all quantities in the solution
as power laws in radius (in particular, it allows us to approximate
$c^2_s = v^2_t/(1 + \beta^{-1})$ as $c^2_s \approx v^2_t(1 -
\beta^{-1})$).  With these assumptions it is relatively
straightforward to obtain the following approximate solution to the
above equations.  At a fiducial radius $r_0$, let $\beta = \beta_0$.
For $r < r_0$, we then have \beq v_t = \sqrt{2 \over 5} r \Omega_K \ \
\ , \ \ \ U_r = - {2 \alpha \over 5} (3/2 - a) r \Omega_K \label{vt},
\eeq \beq \Omega = \Omega_0 \left({ r \over r_0}\right)^{a-3/2} \ \ \
, \ \ \ \beta = \beta_0 \left({r \over r_0}\right)^{-2 a}, \eeq \beq
\Omega_0 = \Omega_K(r_0) \left({(5-3\gt) \over 5 \beta_0 (3/2 - a)
\tilde f} \right)^{1/2}, \eeq \beq \tilde f = 2(\eta + f - 1)/3 +
(1-\eta)(\gt-1),\eeq \beq a = { (5 - 3 \gamma_t) (1 - f - \eta) \over
6(\gamma_t - 1)(\eta - 1) - 4f},\label{a} \eeq and \beq \eta = {(1-f)(5 - 3\gt
-4a) + 6a(\gt -1) \over 5-3\gt - a(4 - 6(\gt-1))}.
\label{eta} \eeq For $a = 0$, $\eta = 1 - f$, and the above solution
reduces to that of NY94.  Equations (\ref{vt})-(\ref{eta}) are
therefore the generalization of the NY94 self-similar solution to the
case when particle heating is explicitly included in a flow with
non-relativistic particles.  We note that, even though the flow has
$v^2_t \propto \rho^{2/3}$, which was the origin of the difficulty in
NY94's self-similar solution, particle heating is accounted for in the
present solution; this is because $\beta \ne$ constant, so that
neither the gas nor the turbulence is strictly virial (although their
sum is).

The key feature of the above solution is that $\beta$ increases as a
function of decreasing $r$ ($a \ge 0$).  In the self-similar solution
of NY94 all of the energy generated by turbulent stresses goes into
maintaining the turbulence in equipartition with the gas (regardless
of the magnitude of $\beta$).  Explicit inclusion of particle heating
therefore must (and does) yield subthermal magnetic field strengths.
Once some of the energy goes into particle heating, there is
insufficient energy available to maintain constant $\beta$ (given the
constraints imposed by being virial).  The increase in $\beta$ is
accompanied by a decrease in $\Omega/\Omega_K$ (recall that the NY94
solution has constant $\Omega/\Omega_K$).  The total pressure and
radial velocity (and hence density), however, have the same (virial)
radial dependence as in NY94.

{\em A priori}, we expect $H \sim q^+$.  We therefore take $\eta =
1/2$ (equal turbulent and particle heating) as a canonical value.  For
$f = 1$, this yields $a = 1/14$ and $\beta \propto r^{-1/7}$ for
$\gamma_t = 1.5$ and $a = 1/6$ and $\beta \propto r^{-1/3}$ for
$\gamma_t = 4/3$ (isotropic magnetically dominated turbulence, for
which $c^2_t = b^2/24\pi\rho$).  For an ADAF with a radial extent
$\sim 10^4$ Schwarzschild radii, $\beta$ changes (over the whole flow)
by a factor $\sim 4$ ($\sim 20$) for $\gamma_t = 1.5 \ (4/3)$.
Explicit inclusion of particle heating can therefore lead to
noticeably subthermal magnetic fields.  The precise numerical factor
by which $\beta$ changes is, however, quite sensitive to the (poorly
understood) choice of $\gamma_t$ and $\eta$.

There are two unsatisfying features of this analytical solution. The
first is the strong dependence on the outer boundary condition; the
radial power law for $\beta$ implies that we must specify its value at
some fiducial radius; this in turn determines the value of $\beta$ in
the interior (where the emissivity peaks).  Secondly, the analytical
solution assumes that, for $f=1$, constant $\beta$ corresponds to
$\eta = 0$ (no heating of particles).  While correct for the
self-similar solution of NY94, this is not strictly correct for global
models of ADAFs (see below).  This motivates us to consider
global solutions of equations (\ref{masscons})-(\ref{angulareq}) and
(\ref{partesimp})-(\ref{tesimp}).  The unsettling features of the
analytical solution are {\em not} reproduced by these numerical
calculations, although the general inference of subthermal magnetic
fields is.

\subsection{Global Dynamics}

We have calculated global (transonic) ADAF models by solving equations
(\ref{masscons})-(\ref{angulareq}) and
(\ref{partesimp})-(\ref{tesimp}) numerically using a relaxation
method.  Here we discuss the numerical method used; in \S4.2.1 we give
sample solutions.  The implications of these calculations are
discussed in \S5.  The models discussed in this section are the
generalization of NKH's global calculations to the case when particle
heating is explicitly included.  

We work in the pseudo-Newtonian potential introduced by Paczy\'nski \&
Wiita (1980), for which $\Omega^2_K = GM/[r(r-r_g)^2]$, where $r_g$ is
the Schwarzschild radius and $M$ is the mass of the black hole.
Integrating equation (\ref{masscons}) implies $1 = - \rho U_r r^2
\Ht$, where we normalize the density by taking the accretion rate to
be $\dot M = 4 \pi$.  Using equation (\ref{masscons}) and the
expression for $\Ht$ from above, we integrate equation
(\ref{angulareq}) once to find \beq {d \Omega \over d R} = {U_r
\Omega_K (\Omega r^2 - j) \over \alpha r^2 v^2_t}, \label{ang} \eeq
where $j$ is the specific angular momentum accreted by the black hole.

Equations (\ref{radialeq}), (\ref{partesimp})-(\ref{tesimp}), and
(\ref{ang}) are the four first order differential equations which we
solve (mass conservation simply normalizes the density).  We also have
two eigenvalues, $j$ and $r_s$, the location of the sonic point.  We
therefore require 6 boundary conditions.

Three of the boundary conditions are the values of $\Omega$, $c_s$,
and $c_t$ (or equivalently, $\Omega$, $v_t$ and $\beta$) at the outer
boundary (which we usually take to be at $r_{out} = 10^5 r_g$).  We
typically specify $\beta$ at the outer boundary and use the
self-similar solution to calculate $\Omega$, $c_s$, and $c_t$.

Equations (\ref{masscons})-(\ref{angulareq}) and
(\ref{partesimp})-(\ref{tesimp}) can, using standard techniques, be
manipulated to show the presence of a sonic point (at $r = r_s$), at
which $ d \ln |U_r|/dr = N/D$, where \beq D = {2p + 2 \over p + 2} -
{U^2_r \over v^2_t} , \eeq \beq N = {r (\Omega_K^2 - \Omega^2) \over
v^2_t} - d\left({1 \over r} - {d \ln \Omega_K \over dr}\right) +
{{\tilde f} q^+ \over \rho U_r v^2_t (p + 2)} \label{num}, \eeq \beq
\tilde f = (\gamma - 1)(\eta + f - 1) + (\gt - 1)(\eta - 1), \eeq and
\beq p = {(\gamma - 1)c^2_s + (\gt - 1)c^2_t \over v_t^2}. \eeq A
well-behaved solution must have $N = D = 0$ at $r = r_s$, which
provides two additional boundary conditions.  The location of the
sonic point is not predetermined but is our second eigenvalue (it is
typically at $\sim 5 r_g$ for $\alpha \sim 0.3$).

The final boundary condition is the torque free condition, $j = \Omega
r^2$, at the horizon.  For numerical reasons, we typically apply this
condition at $r = 1.5 r_g$, but we have confirmed that the results are
unchanged if we apply it at $r = 1.1 r_g$.  Applying this final
boundary condition at the horizon is a defect of our acausal treatment
of viscosity (see NKH for a discussion of this point and Gammie \&
Popham 1998 for a causal treatment of viscosity in ADAF models).

\subsubsection{Results}

For the numerical results in this section we set $f = 1$, $\gamma =
5/3$, $\gamma_t = 1.5$, and $\alpha = 0.3$, and explore the solutions
as a function of $\eta$ and the value of $\beta$ at the outer
boundary.  We focus on understanding the evolution of $\beta$ as a
function of $r$, and the relationship between $\beta$ and $\eta$.

To start with, we apply the self-similar boundary condition
corresponding to $\beta = 1$ at $r = r_{out}$.  Figure 1a shows the
radial velocity ($U_r$) and total sound speed ($v_t$) for models
ranging from $\eta = 0$ to $\eta = 1$.  Figure 1b shows the density
for the same models.  As predicted by the self-similar solution, the
basic dynamical quantities (total pressure, density, etc.)  are
independent of the details of the particle heating (i.e., $\eta$).
The insensitivity of $\rho$, $v_t$, and $U_r$ to $\eta$ is a generic
feature of our models; we therefore do not show additional plots of
these quantities.

Figures 1c and 1d show $\Omega/\Omega_K$ and $\beta$ as functions of
$r$ for $\eta$ varying from $0$ to $1$ in steps of 0.2.  Here there is
a nontrivial dependence on $\eta$.  In order to understand the
behavior of $\beta(r)$, the solid lines in Figure 2a show $\beta(r)$
for an alternative outer boundary condition, $\beta = 15$ at $r =
r_{out}$ (all other parameters are the same as in Figure 1).

The self-similar solution of \S4.1 predicts that $\beta$ should be a
monotonically increasing function of $r$, namely, $\beta \propto
\beta_{out} r^{-2a}$ where, for a fixed $\gt$, $a \ge 0$ is just a
function of $\eta$ (see eq. [\ref{a}]).  For $\eta \ne 1$, this is not
always seen in the global calculations.  In fact, in contrast to the
self-similar solution, $\beta$ shows a tendency to approach a roughly
constant value at small $r$ (especially near the sonic point, inside
of which the fluid flows in nearly adiabatically), and to decrease
with radius if it starts off with a large value
(Fig. 2).

Deviations from self-similarity in the density and temperature
profiles of the flow, driven primarily by the existence of a sonic
point, which modifies the radial velocity and therefore $\rho$, can
lead to more noticeable changes in the entropy gradient (because the
entropy gradient involves a cancellation between the temperature and
density gradients).  Since the value of $\eta$ is $\propto$ the
entropy gradient (cf eqs. [\ref{partesimp}]-[\ref{eta2}]), it, and
hence $\beta$, is sensitive to such deviations.  This is the basic
reason why self-similarity is less accurate when calculating $\eta$
and $\beta$, even though it is reasonably good for $\rho$, $c^2_s$,
etc..

Nonetheless, the key qualitative conclusion of the self-similar
solution is confirmed by the global calculations. Explicit inclusion
of particle heating at reasonable levels ($\eta \sim 1/2$) leads to
subthermal magnetic fields.  In fact, there is an approximate mapping
between the value of $\beta$ in the interior of the flow and $\eta$,
which is independent of the outer boundary condition (this is in
contrast to the self-similar solution, which was a power law in $r$,
and thus very sensitive to the outer boundary condition).  This is
seen in Figure 2b, which shows $\beta(r)$ for $\eta = 1/2$, taking
$\beta_{out} = 1,5,15,$ and $50$.  In the inner portions of the flow,
where the emissivity peaks, the solutions roughly converge to a common
$\beta \sim 5$.

Furthermore, the global calculations confirm that strong levels of
turbulence in the interior of the flow are incompatible with
reasonable levels of particle heating.  Strictly equipartition fields
($\beta = 1$) near $r \sim 1$ are only obtained for unphysically small
values of $\eta \lsim 0.1$ (Fig. 1d and 2a).

\section{Discussion}

In \S2 of this paper we derived the energy equations for particles,
magnetic fields, and turbulence in optically thin, two temperature
ADAFs (using mean field MHD).  The fundamental energy equations are
given in equations (\ref{ben}) - (\ref{stress}); for explicit
calculations, however, we have used drastically simplified versions
which approximate the turbulence as an additional fluid in the problem
(the approximations required are detailed in \S2.2).  One application
of this analysis, given in  Appendix A, is to resolve a disagreement
in the literature between several treatments of the energetics of
ADAFs.

The primary focus of this paper, however, has been to clarify and
explore a previously underappreciated piece of physics regarding
optically thin, two-temperature ADAFs.  The issues addressed stem from
the fact that non-relativistic ADAFs are, at some level, a singular
problem.  Strong advection leads to a virial flow in which
non-relativistic particles tend to be isentropic.  This is because, at
least in the self-similar regime, being virial requires $T \propto
\rho^{2/3}$, which is an adiabat for non-relativistic particles with
$\gamma = 5/3$.  On the other hand, the turbulent heating of particles
should cause the entropy of the gas to increase inwards.  To resolve
these conflicting requirements, either the flow must be non-rotating
(unlikely) or else a formalism must be developed which allows for both
differential rotation and particle heating.

This issue is particularly relevant for optically thin, two
temperature, ADAFs which have been extensively applied to observed
systems; for such flows, $T_i \gg T_e$ and the gas is quite accurately
approximated as non-relativistic.  Following NY95, a number of workers
have suggested that, because of the ``singularity,'' a proper
treatment of the flow dynamics should include the turbulent/magnetic
contribution to the pressure and energy density (the electron
contribution is expected to be less important if $\beta \lsim
T_i/T_e$, which is well satisfied in most models).  The derivation in
\S2 ``formalizes'' this suggestion.

Previous attempts at phenomenologically including turbulence in the
dynamics of ADAFs assumed a constant $\beta$ for the flow (e.g., NY95,
EMN, Narayan et al. 1998a).  What was not appreciated, however, is
that this still does not account for the heating of particles; rather,
the implicit assumption is that most of the energy generated by shear
in the flow is stored as turbulent energy (\S3).  This energy is in
fact precisely what is required to maintain the constant $\beta$.

The scenario required by these models is unlikely to be realized in
real flows.  Strongly turbulent plasmas inevitably heat particles
(energy cascades to small scales, etc.).  To assess the implications
of this, we impose particle heating in the dynamical equations by
specifying that some fraction ($\equiv \eta$) of the energy generated
by turbulent stresses heats the particles.  The remaining energy (a
fraction $1 - \eta$) is used to build up the magnetic field strength
and turbulent kinetic energy as the gas flows in.  The parameter
$\eta$ therefore replaces the parameter $\beta$.  The
turbulence/magnetic field in the flow ($\beta$) is now an output of
the dynamical equations; it evolves so as to maintain the specified
level of particle heating.

For thin accretion disks, one expects $\eta = 1$ to good approximation
due to the mismatch between the timescale on which the gas pressure
increases (the inflow time) and the timescale on which shear acts to
increase the turbulent energy (the rotational period).  For $\alpha$
not $\ll 1$, no such mismatch exists for ADAFs and so $\eta \ne 1$,
although we do expect $\eta \sim 1$ (rather than $\eta \ll 1$; \S2.3).

We have given analytical and numerical solutions for the dynamics of
ADAFs at constant $\eta$.  These are, respectively, the generalization
of the NY94 self-similar solution and NKH and Chen et al.'s (1997)
global calculations (in the Paczy\'nski potential) to the case where
particle heating is explicitly included.  The results of the
self-similar solution are particularly straightforward to understand.
In previous, constant $\beta$, models, the constancy of $\beta$ was
achieved at the expense of particle heating.  If, instead, we require
a reasonable level of particle heating, less energy is available for
the turbulence and so $\beta$ must increase as the gas flows in. The
turbulence/magnetic field therefore become subthermal.  We emphasize,
however, that the predicted density, radial velocity, and total
pressure of the flow are relatively unchanged (with respect to NY94 or
NKH) by the present analysis (\S4; Figure 1).

Self-similarity requires that $\beta$ monotonically increase as the
gas flows in, while our global calculations show that this is not
reproduced if the turbulence is somewhat subthermal.  The global
calculations do confirm, however, that strong levels of turbulence
($\beta \sim 1$) are incompatible with reasonable levels of particle
heating (since they require $\eta \ll 1$; see Figures 1d and 2a).  If
we prioritize particle heating, taking $\eta \sim 1/2$, we find that
$\beta$ converges to $\sim 5$ in the interior of the flow, regardless
of its initial value at large radii (Figure 2b).

Explicitly calculating $\beta$ requires understanding the non-linear
saturation of the turbulence; this would provide the full profile,
$\eta(r)$.  Our point is that, regardless of the details of the
non-linear saturation, one expects that $\eta$ is not $\ll 1$.  Taking
a lower bound of $\eta \sim 1/2$ yields the corresponding bound,
$\beta \gsim 5$.

The $\beta$ used in this paper refers to both the magnetic energy in
the turbulence and the turbulent kinetic energy (cf eq. [\ref{beta}]);
in particular, we assume roughly Alfvenic turbulence, for which these
two energies are equal.  Our analysis suggests that consistency with
particle heating requires both subthermal magnetic and turbulent
kinetic energies; on dimensional grounds, i.e., independent of the
physics of angular momentum transport, this requires a relatively
smaller $\alpha$.  This can be seen by noting that $\alpha \approx -
T_{r \phi}/\rho c^2_s$ in a Shakura \& Sunyaev-like closure relation,
where $c_s$ is the sound speed.  From the definition of the stress
tensor (eq. [\ref{stress}]), we expect $T_{r \phi} \lsim \rho c^2_t$,
where $c^2_t = c^2_s/\beta$ is the square of the turbulent velocity.
Dimensional analysis therefore implies $\alpha \lsim 1/\beta$.
Quantitatively, then, the constraint $\beta \gsim 5$ suggests that
$\alpha \lsim 0.2$.

\subsection{Caveats}

To conclude, we discuss several key assumptions implicit (or explicit)
in our analysis.

First, the correct value for $\eta(r)$ is highly uncertain, and the
inferred $\beta$ depends non-trivially on the $\eta$ chosen (see
Figures 1 \& 2).  For example, one might think that $\eta$ and $\beta$
would be anti-correlated (i.e., preferential build up of the
turbulence if it is subthermal).  In principle, one could solve the
mean field energy equations (eqs. [\ref{ben}] - [\ref{pen}]) using,
e.g., a second order closure scheme, and thus calculate $\eta$
(Speziale 1991), but this is well beyond the scope of our paper.

Second, we treat the turbulence/magnetic fields in the flow using a
crude fluid model.  Even within the context of this model, we assume
that the effect of flux freezing/compression on the turbulent energy
is insufficient to bring the turbulence into equipartition with the
gas ($\gamma_t \lsim 5/3$).  This requires that less than half of the
magnetic energy is in radial perturbations and that more than a third
of the turbulent kinetic energy is in radial perturbations (Appendix
B).  Our physical picture is that the primary source of the turbulent
energy is shear in the mean flow.

A noticeable omission is our neglect of third order (and higher)
turbulent quantities (\S2.2).  These represent, among other things,
the turbulent diffusion of energy/entropy, and thus are quite relevant
for the issues discussed in this paper.  Unfortunately, turbulent
transport is particularly complicated in ADAFs, since both convection
and MHD turbulence are believed to be important.

Although we consider a radially varying $\beta$, we take $\alpha$ to
be a constant.  Aside from simplicity, our motivation for this
restriction is that there is generally no self-similar
advection-dominated accretion solution when $\alpha$ decreases as the
gas flows in (which would be expected by our analysis if $\alpha
\propto \beta^{-1}$).  This is because a decreasing $\alpha$ leads to
an increase in density more rapid than $r^{-3/2}$ (if $\alpha \propto
r^{g}$, $U_r \propto r^{-1/2 + g}$, and $\rho \propto r^{-3/2 -g}$).
The compressive heating associated with this density profile is
sufficiently large that it generally yields a {\em decreasing} entropy
gradient for the flow.  The only viable self-similar solution in this
case is one which is {\em cooling} dominated, rather than
advection-dominated.

Finally, our calculations are (as is standard) a one dimensional
(height integrated) analysis of a three dimensional problem.  Relaxing
this restriction will likely make the singularity inherent in the
dynamical equations less severe.  The extent to which this will
alleviate our concerns about particle heating is, of course, unclear.

\noindent{\it Acknowledgments.}  It is a pleasure to thank Charles
Gammie and Andrei Gruzinov for useful discussions. EQ was supported
by an NSF Graduate Research Fellowship; this work was also supported
by NSF Grant AST 9423209.

\newpage

\begin{appendix}

\section{Related Work}
Here we compare our analysis of the energetics of ADAFs with several
other treatments in the literature, and clarify several issues about
the use of ``effective'' adiabatic indices.

EMN treat the energetics of an ADAF using both a total energy equation
and an electron energy equation.  Our derivation/discussion of the
total energy equation in \S2 is a more rigorous treatment of the
problem, which arrives at conclusions similar to theirs (which is
based on Esin 1997).  The one difference is that they neglect the
turbulent velocity (i.e., $\beta_v \rightarrow \infty$), so that their
expressions for $c_t^2$ and $\gamma_g$ are slightly different (the
most general expression for $\gamma_g$ is given by eq. [\ref{gtotgen}]
of Appendix B).

EMN's treatment of the electron energy equation (their Appendix A) is,
however, quite different from ours.  They take the electron internal
energy and pressure to be effective quantities, containing
contributions from the magnetic field.  They thus argue that the
adiabatic index which shows up in the electron energy equation is also
an effective $\gamma$ (in fact they take $\gamma_e = \gamma_g$). The
physical picture on which this is based is that (1) {\em all} of the
energy generated by turbulent stresses heats the particles, i.e., $H =
q^+$, and (2) the energy required to bring the magnetic energy from
its flux-frozen value to the equipartition value comes from the
thermal energy of the particles (hence $\gamma_e$ smaller than its
monatomic value).  The problem with this analysis is that it
incorrectly treats the thermal energy of the particles as a source of
magnetic energy.  Within the context of MHD, turbulence and magnetic
fields can act as a source of thermal energy, while the converse is
not possible (this can be seen formally using the mean field energy
equations in \S2).  If we set $H_e = 0$ (no turbulent heating of
electrons), then the turbulence and magnetic field should not appear
in the electron energy equation (except as a third order turbulent
diffusion term), while in EMN's treatment they still would (through
the effective $\gamma$).  To conclude, then, the appropriate adiabatic
index in the electron energy equation is not an ``effective'' one, but
rather that of a monatomic ideal gas.

Manmoto et al. (1997) and Nakamura et al. (1997) also solve the energy
equations for ADAFs.  They take electron and ion equations analogous
to (\ref{pe}) and (\ref{ee}) as their basic energy equations, with
$H_e = 0$, $H_i = q^+$, and $\gamma_e = \gamma_i = 5/3$.  Setting $H_e
= 0$ assumes no turbulent heating of electrons, which we regard as
overly optimistic.  Setting $\gamma_e = 5/3$ is correct in spirit (in
that there is no ``turbulent'' component to $\gamma_e$), but it
assumes non-relativistic electrons; this is incorrect in the inner
$\sim 10^2-10^3$ Schwarzschild radii.  Setting $H_i = q^+$ assumes
that all of the energy generated by turbulent stresses heats the ions.
As emphasized in this paper, however, $H$ is, in general, $\ne q^+$
(although it is likely $\sim q^+$).

There is a close relationship between the ``ion'' and ``total'' energy
equations discussed above and in \S2.  In particular, in treating the
``ion/total'' energy equation, one can either use a monoatomic
adiabatic index {\em or} an effective adiabatic index.  If the former,
the heating term should be $H \ne q^+$, since one is truly writing a
particle energy equation.  If the latter, the heating term should be
$q^+$ (see eq. [\ref{esimple}]). Put another way, it is precisely the
inequality of $H$ and $q^+$ which allows one to use an effective
adiabatic index.  This can be seen explicitly by ``solving'' the
turbulent energy equation (eq. [\ref{turbesimp}]) for $H$ and
substituting it into the particle energy equation (eq. [\ref{pen}]),
in which case one readily derives the total energy equation
(eq. [\ref{esimple}]).

No such analysis is allowed for the electron energy equation, however,
since it is a standard assumption of ADAF models that $T_e \ll T_i$
and the electrons couple poorly to the turbulence ($\delta_H \ll 1$).
The uncertainty introduced by $H \ne q^+$ is therefore absorbed into
an uncertainty in (the already uncertain) electron heating parameter,
$\delta_H$, rather than in the adiabatic index.

The above arguments regarding the electron adiabatic index in ADAFs
break down if the distribution functions are anisotropic.  In this
case adiabaticity cannot be described by a single adiabatic index (cf
double adiabatic theory).  In particular, for anisotropic distribution
functions, thermal energy can be a source of magnetic energy (the
fire-hose instability being a well known example), contrary to the
arguments using MHD given above.  ADAFs are nearly collisionless
systems; isotropy is therefore not guaranteed.  Nonetheless, we (as
with all other workers in the field) have proceeded on the assumption
that it is a reasonable approximation.  Anisotropic distribution
functions are known to be unstable to generating modes (e.g., Alfven
waves, fast modes, whistlers) which act to restore isotropy (e.g.,
Melrose 1980).  The timescale for such processes is typically of order
the cyclotron period, much shorter than the dynamical time in the
accretion flow.

\section{Anisotropic Turbulence}

When simplifying the mean field energy equations in \S2, we assumed
that the diagonal components of the turbulent stress tensor are all
equal (i.e., $b^2_r \approx b^2_\phi \approx b^2_\theta$, etc.).  This
is equivalent to assuming that the ``turbulent pressure'' is
isotropic.  In this Appendix, we generalize the derivation of \S2 to
arbitrary diagonal components for the stress tensor.

Assuming azimuthal symmetry, $U_\theta = 0$, and uniform rotation on
spherical shells, but allowing for arbitrary values of $T_{rr},
T_{\theta \theta},$ and $T_{\phi \phi}$, the energy generation term of
equation (\ref{totg}), which shows up in the total energy equation for
the fluid, becomes
\beq G = - \left( \left \langle \ve
\right \rangle + {2} \left \langle \be \right \rangle + \gamma
\langle \epsilon\rangle \right) \partial_k U_k - T_{rr} {d U_r \over d r}
- (T_{\theta \theta} + T_{\phi \phi}) {U_r \over r} - G_{od},
\label{ggen} \eeq where 
\beq G_{od} = T_{r \phi} \sin \theta r {d \Omega \over d r} -
T_{\theta \phi} \Omega \cos \theta \eeq is the contribution to $G$
from the off-diagonal components of the stress tensor.  This is the
same as derived in \S2.2.

For $T_{rr} = T_{\theta \theta} = T_{\phi \phi}$, as taken in \S2,
equation (\ref{ggen}) reduces to equation (\ref{gsimp}).  This is
because $dU_r/dr + 2U_r/r = \nabla \cdot {\bf U}$ under the
assumptions of $U_\theta = 0$ and azimuthal symmetry.  In this case,
the terms associated with geometrical convergence of the flow in
spherical coordinates (the $dU_r/dr$ and $U_r/r$ terms) are equivalent
to a term $\propto d \rho/ dr$, i.e., a $PdV$ work term.

As equation (\ref{ggen}) shows, however, one cannot in general write
the ``geometrical'' terms as ``compressive'' terms and thus one cannot
derive a fluid approximation for the turbulence in the flow.  Assuming
self-similarity, $U_r \propto r^{-1/2}$, allows one to do so.  In this
case, $dU_r/dr \approx - \nabla \cdot {\bf U}/3$ and $U_r/r \approx 2
\nabla \cdot {\bf U}/3$.  With this simplification, equation
(\ref{ggen}) reduces to

\beq G = - \left[ \left({7 \over 3} - 2 h_v \right) \left \langle \ve
\right \rangle + {\left({2 \over 3} + 2 h_b\right)} \left \langle \be
\right \rangle + \gamma \langle \epsilon\rangle \right] \partial_k U_k
- G_{od},
\label{ggensimp} \eeq where \beq h_v = {\langle v_r^2 \rangle \over
\langle v^2 \rangle} \ \ {\rm and} \ \ h_b = {\langle b_r^2 \rangle
\over \langle b^2 \rangle} \eeq are the fraction of the turbulent
kinetic energy ($h_v$) and magnetic energy ($h_b$) in radial
perturbations, respectively.

Expressing the level of turbulence in the flow in terms of the
$\beta$'s of equation (\ref{beta}) allows one to derive a total energy
equation for the flow which is identical in form to equation
(\ref{esimple}).  The expressions for the total sound speed, $v_t$,
and the effective adiabatic index, $\gamma_g$, are, however,
different.  They are now given by

\beq v^2_t = c^2_s \left[1 + \beta_v^{-1}\left({4 \over 3} - 2
h_v\right) + \beta^{-1}_b\left(2h_b - {1 \over 3}\right)\right]
\label{vtgen} \eeq and 

\beq \gamma_g = {\gamma +
\left(\gamma-1\right)\beta^{-1}_v\left({7\over3} - 2 h_v\right) +
\left(\gamma-1\right)\beta^{-1}_b\left({2\over3} + 2 h_b\right) \over
1 + \left(\gamma-1\right)\beta^{-1}_v +
\left(\gamma-1\right)\beta^{-1}_b}. \label{gtotgen} \eeq If we set
$h_v = h_b = 1/3$ and $\beta_b = \beta_v \equiv \beta$, equations
(\ref{vtgen}) and (\ref{gtotgen}) reduce to equations (\ref{speed})
and (\ref{gamma}), respectively.

Equation (\ref{gtotgen}) demonstrates the simple, but important, point
that the effect of flux freezing/compression on the turbulent energy
depends on the {\em geometry} of the turbulence.  If we set $\beta_b
\rightarrow 0$ in equation (\ref{gtotgen}), we recover the ``adiabatic
index'' associated with flux freezing of the magnetic field, namely
\beq \gamma_b = {2 \over 3} + 2 h_b. \label{gb} \eeq The corresponding
adiabatic index for the velocity component of the turbulence is \beq
\gamma_v = {7 \over 3} - 2h_v. \label{gv} \eeq It should be emphasized
again that equations (\ref{ggensimp})-(\ref{gv}) are strictly valid only
in the self-similar regime.

In the bulk of this paper we have assumed that flux freezing is
insufficient to keep the magnetic field/turbulence in equipartition
with the gas, i.e., that $\gamma_b \lsim 5/3$.  By equation
(\ref{gb}), this can be recast as an assumption on the magnetic field
geometry, namely $h_b \lsim 1/2$, i.e., less than half of the magnetic
energy is in radial perturbations.  The corresponding constraint on
the turbulent velocity field is that $h_v \gsim 1/3$, i.e., more than
$1/3$ of the turbulent kinetic energy is in radial perturbations.

Both of these constraints ($h_b \lsim 1/2$ and $h_v \gsim 1/3$) are
satisfied in numerical simulations of thin accretion disks (cf Hawley,
Gammie, \& Balbus 1996).  These are, however, of questionable
relevance for the accretion flows at hand. For laminar, non-rotating,
spherical accretion, it is known that $h_b \sim 1$: the magnetic field
becomes purely radial (Shapiro 1973), and flux freezing is
energetically important, invalidating the analysis of this paper.
This result is also of questionable relevance for the accretion flows
at hand since, while quasi-spherical, ADAFs are differentially
rotating, turbulent, etc..  We suspect that ADAFs lie somewhere
between the two extremes of thin disks and spherical accretion, but
numerical simulations will be necessary to determine precisely where.

\end{appendix}

\newpage

{
\footnotesize
\StartRef
\noindent {\large \bf References} \\
\Ref Abramowicz, M., Chen, X., Granath, M., \& Lasota, J.-P., 1996, ApJ, 479, 
179 \\
\Ref Abramowicz, M., Chen, X., Kato, S., Lasota, J.-P., \& Regev, O., 1995,
ApJ, 438, L37 \\
\Ref Abramowicz, M., Lanza, A. \& Percival, M. J., 1997, ApJ, 479, 179 \\
\Ref Balbus, S. A., \& Hawley, J.F., 1998, Rev. Mod. Phys., 70, 1 \\
\Ref Bisnovatyi-Kogan, G. S., \& Lovelace R. V. E., 1997, ApJ, 486, L43 \\
\Ref Chen, X., Abramowicz, M.A., Lasota, J.-P., Narayan, R., \& Yi, I. 1995, ApJ, 443, L61 \\
\Ref Chen, X., Abramowicz, M.A., Lasota, J.-P., 1997, ApJ, 476, 61 \\
\Ref Esin, A. A., 1997, ApJ, 482, 400 \\
\Ref Esin, A. A., McClintock, J. E., \& Narayan, R., 1997, ApJ, 489, 86 (EMN)\\
\Ref Gammie, C.F \&  Popham, R.G., 1998, ApJ, 498, 313 \\
\Ref Hawley, J. F., Gammie, C. F., \& Balbus, S. A., 1996, ApJ, 464, 690 \\
\Ref Ichimaru, S. 1977, ApJ, 214, 840 \\
\Ref Kato, S., Abramowicz, M., \& Chen, X. 1996, PASJ, 48, 67 \\
\Ref Kato, S., Fukue, J., Mineshige, S., 1998, {\em Black-Hole Accretion Disks}
(Japan: Kyoto University Press) \\
\Ref Manmoto, T., Mineshige, S., \& Kusunose, M., 1997, ApJ, 489, 791\\
\Ref Melrose, D.B. 1980, {\em Plasma Astrophysics} (New York: Gordon \& Breach) \\
\Ref Nakamura, K. E., Masaaki, K., Matsumoto, R., \& Kato, S. 1997, PASJ, 49, 503 \\
\Ref Narayan, R., Barret, D., \& McClintock, J. E., 1997, ApJ, 482, 448\\
\Ref Narayan, R., Kato, S., \& Honma, F. 1997, ApJ, 476, 49 (NKH)\\
\Ref Narayan, R., \& Yi, I., 1994, ApJ, 428, L13 (NY94)\\ 
\Ref Narayan, R., \& Yi, I., 1995a, ApJ, 444, 231 \\ 
\Ref Narayan, R., \& Yi, I., 1995b, ApJ, 452, 710 (NY95)\\ 
\Ref Narayan, R., Mahadevan, R., Grindlay, J.E., Popham, R.G., \& Gammie, C., 1998a, ApJ, 492, 554 \\
\Ref Narayan, R., Mahadevan, R., \& Quataert, E., 1998b, in {\em The Theory
of Black Hole Accretion Discs}, eds. M.A. Abramowicz, G. Bjornsson, and J.E. Pringle (Cambridge:  Cambridge University Press) (astro-ph/9803131) \\
\Ref Paczy\'nski, B. \& Wiita, P. J., 1980, A\&A, 88, 23 \\
\Ref Parker E.N. 1979, {\em Cosmical Magnetic Fields} (New York: Oxford University Press) \\
\Ref Rees, M. J., Begelman, M. C., Blandford, R. D., \& Phinney,
E. S., 1982, Nature, 295, 17 \\
\Ref Shakura, N. I., \& Sunyaev, R. A., 1973, A\&A, 24, 337 \\
\Ref Shapiro, S. L., 1973, ApJ, 185, 69 \\
\Ref Shapiro, S. L., Lightman, A. P., \& Eardley, D. M. 1976, ApJ, 204, 187 \\ 
\Ref Speziale, C. G., 1991, Ann. Rev. Fluid Mech., 23, 107 \\
}

\newpage  


\begin{figure}
\plottwo{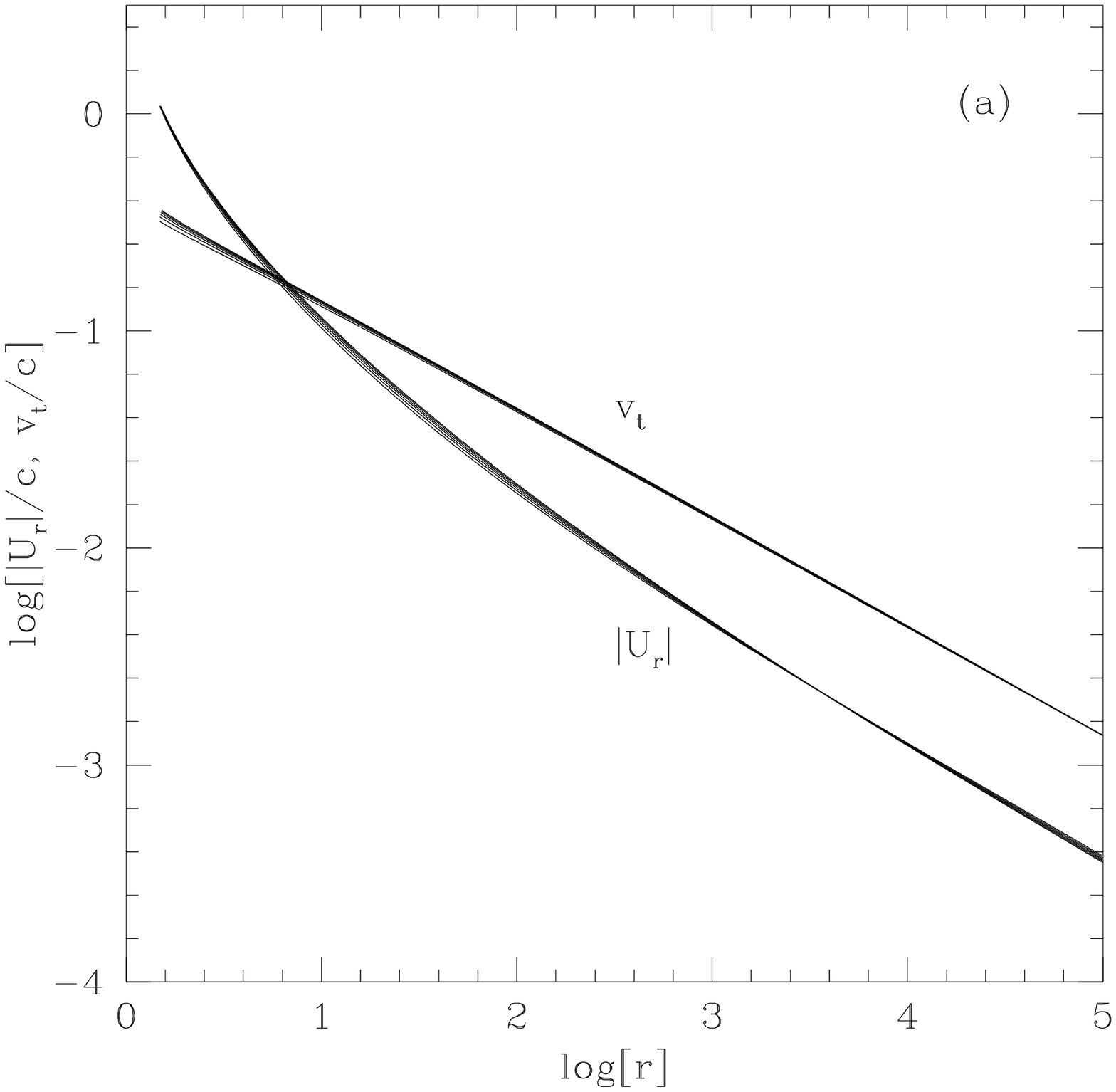}{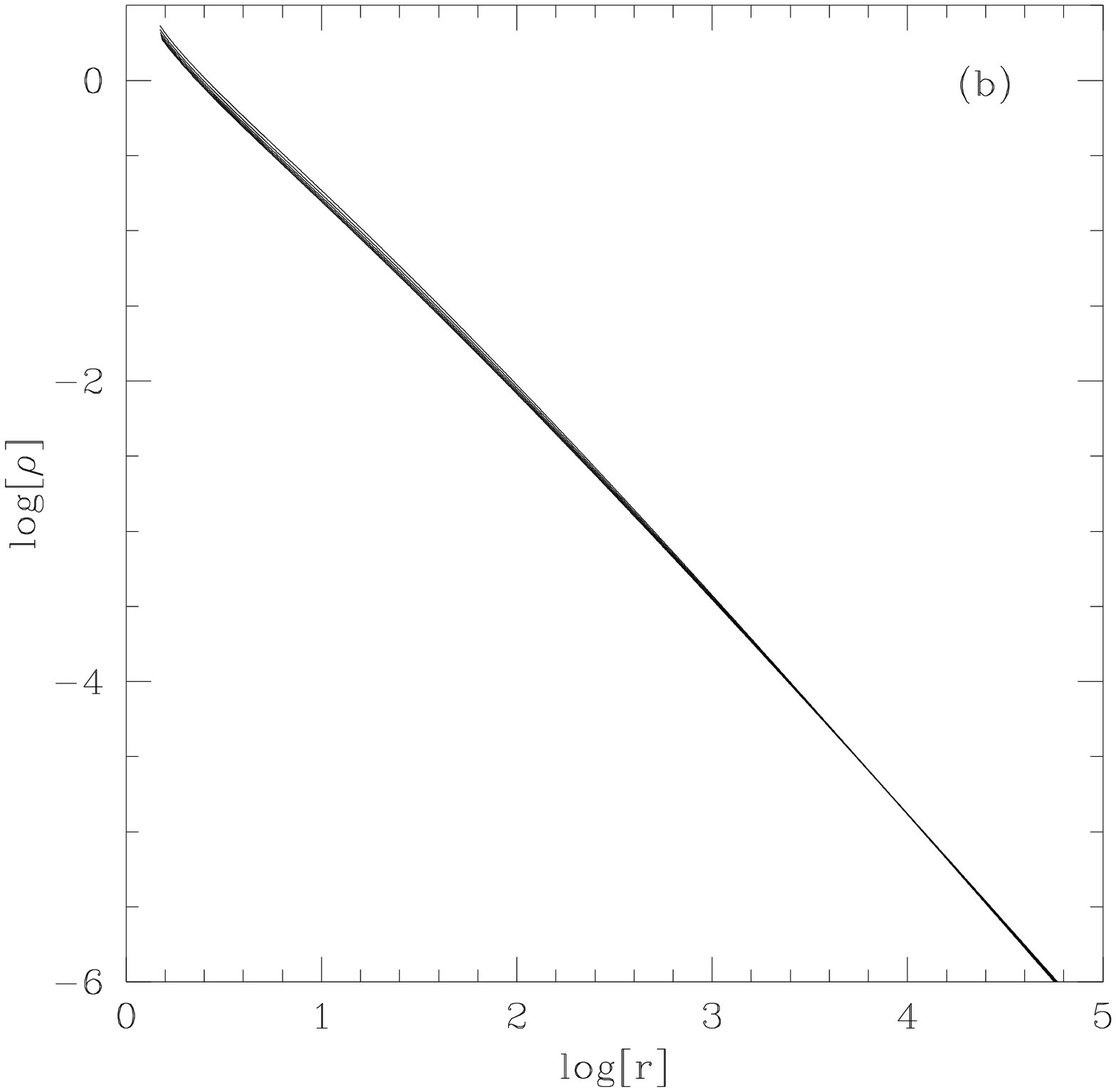} 
\plottwo{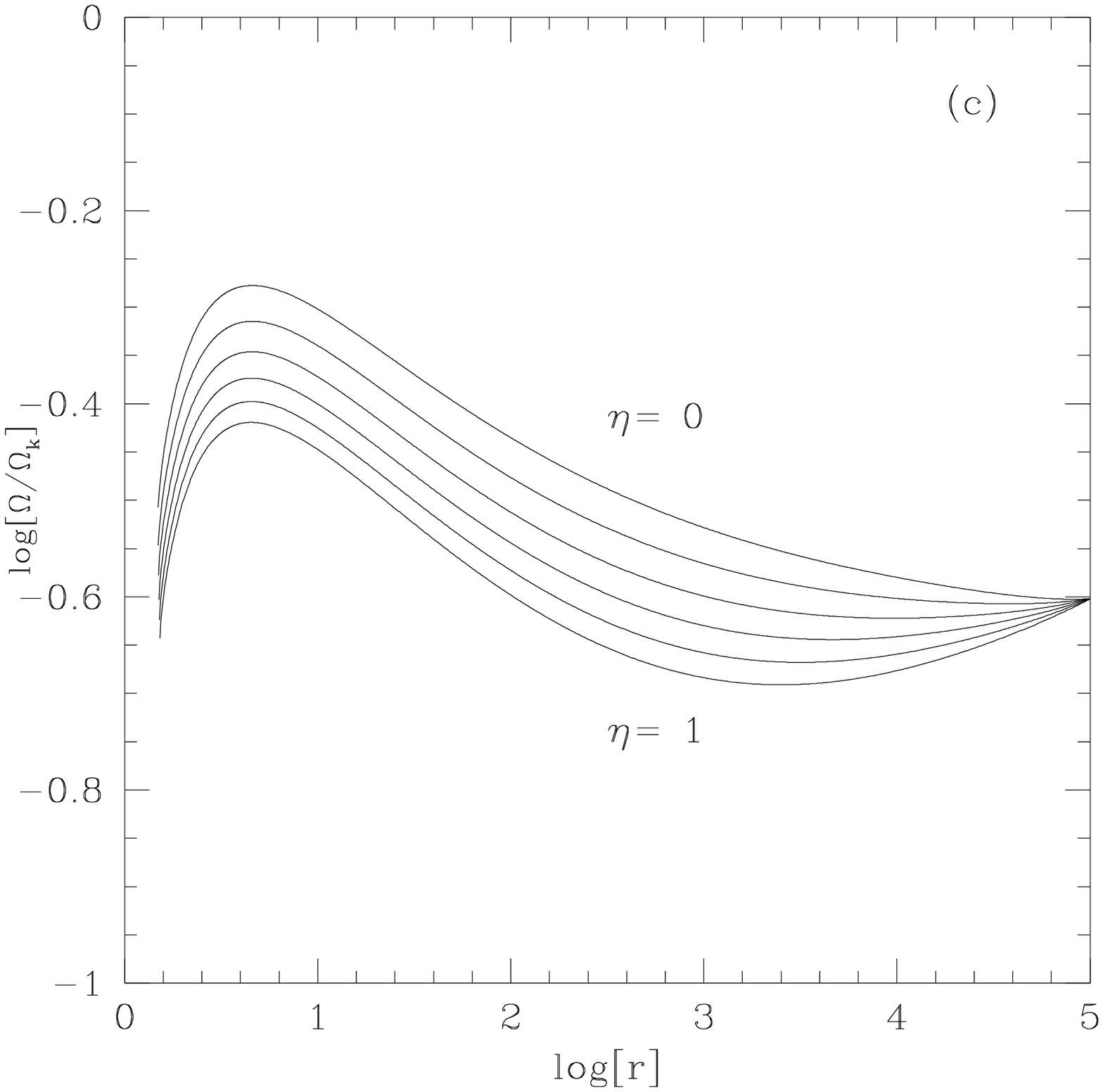}{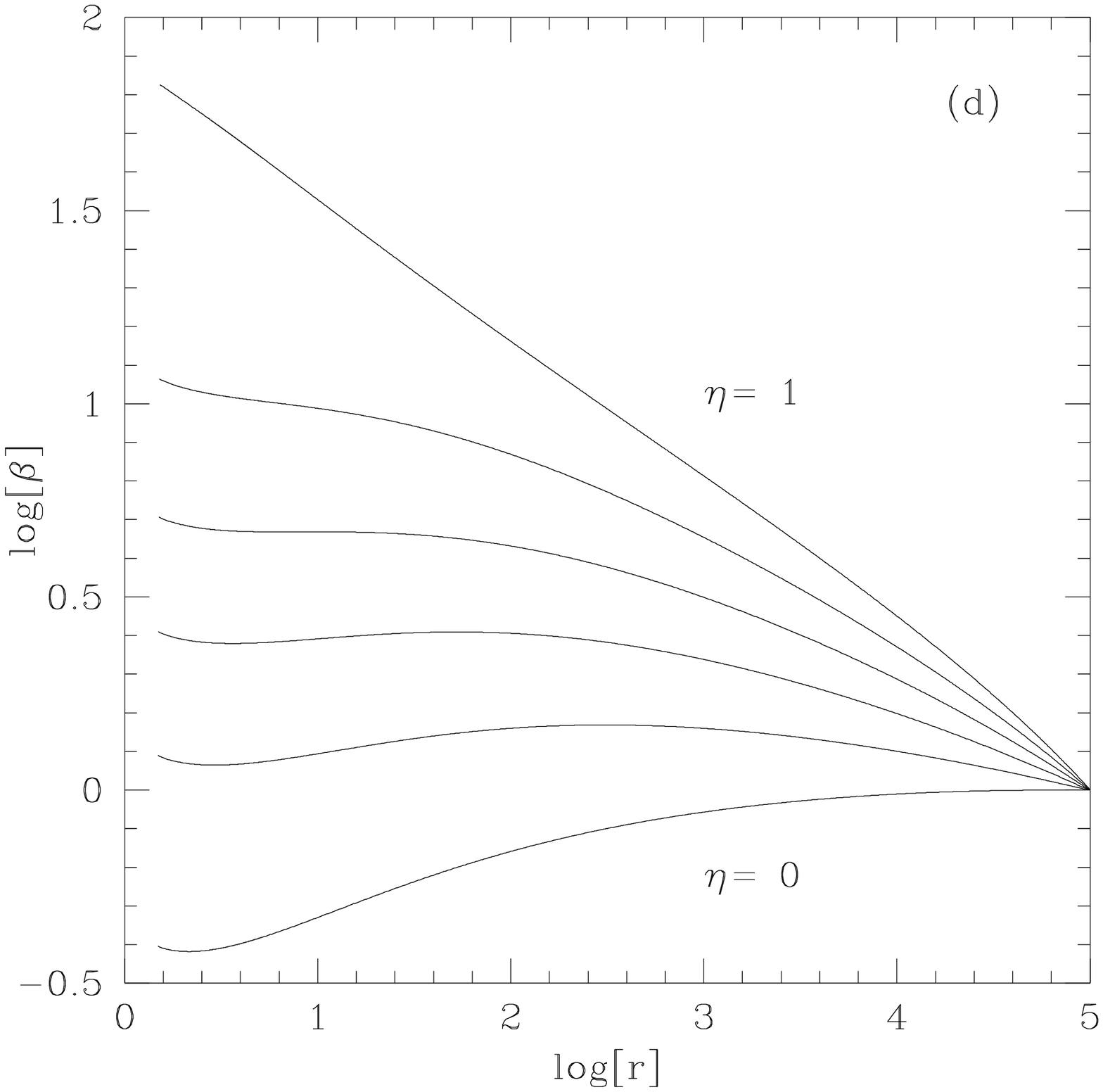}
\caption{Radial profiles of several quantities in our global dynamical
models of ADAFs.  Each panel shows plots for $\eta$ from $0$ to 1 in
steps of 0.2.  Other parameters take the values: $\gamma_t = 1.5,
\alpha = 0.3, f = 1$, and $\beta = 1$ at $r = r_{out} = 10^5 r_g$.
(a) radial velocity ($U_r$) and total sound speed ($v_t$), (b)
density, (c) rotation rate in units of the Keplerian rate, (d) ratio
of gas to turbulent pressure ($\beta$).  Note that $U_r$, $v_t$, and
$\rho$ are essentially independent of $\eta$, while $\Omega/\Omega_K$
and, in particular, $\beta$, depend non-trivially on $\eta$.}
\end{figure}

\newpage

\begin{figure}
\plottwo{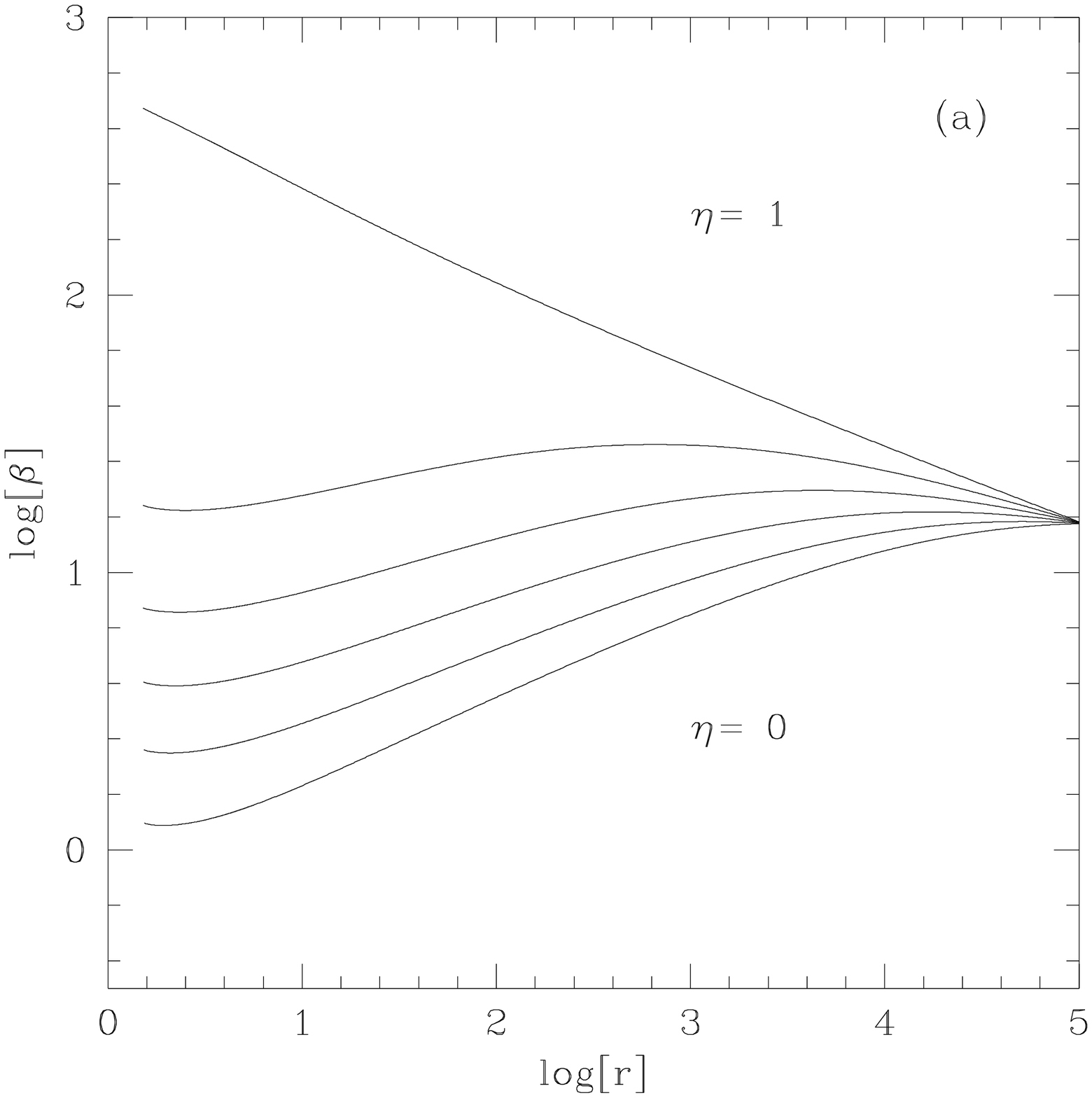}{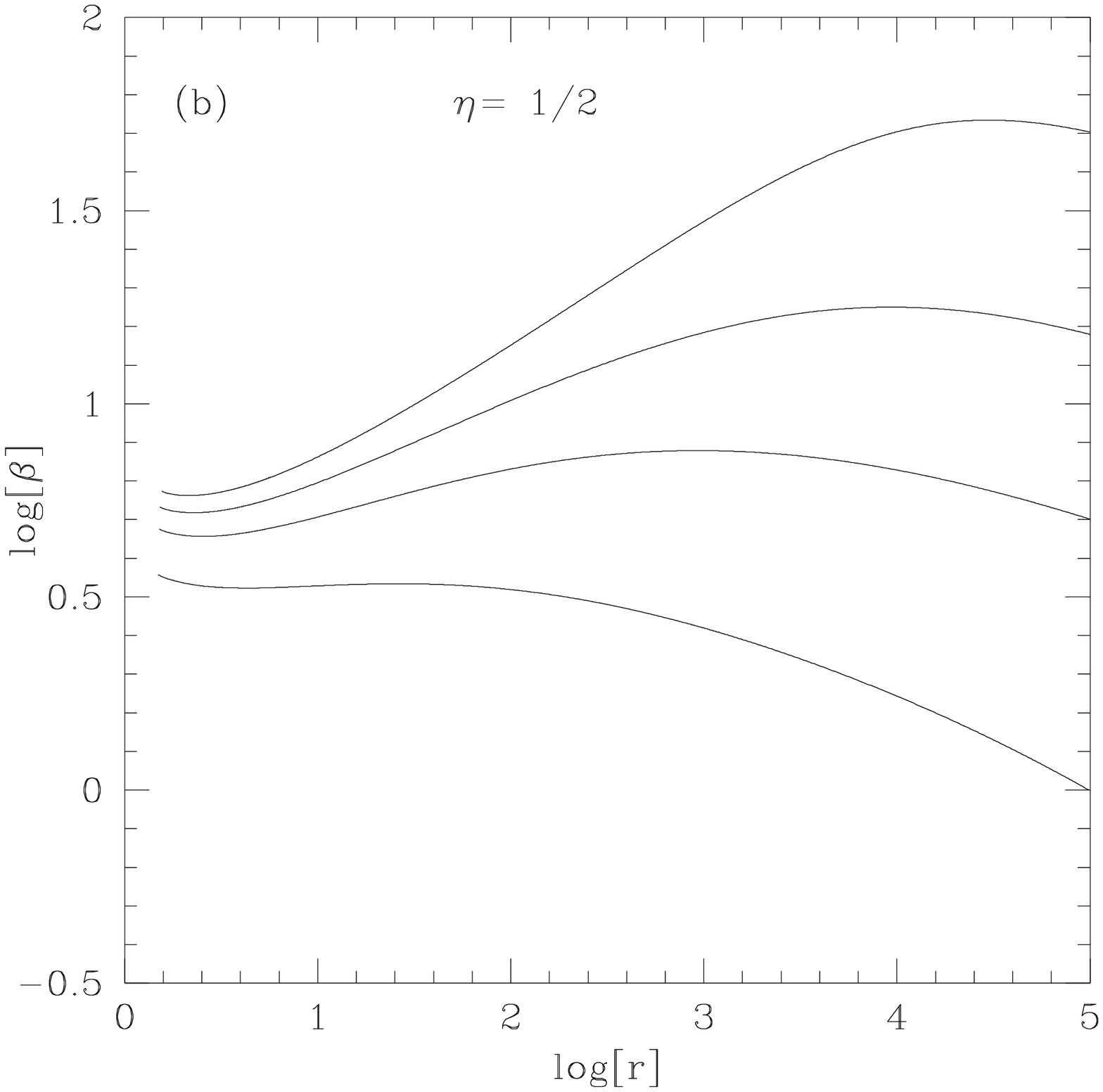}
\caption{(a) $\beta(r)$ for the same parameters as in Figure 1, except
$\beta$ is set to 15 at the outer boundary.  (b) $\beta(r)$ for $\eta
= 1/2$, $\gamma_t = 1.5$, and several different values of $\beta$ at
the outer boundary.  The solutions roughly converge to a common $\beta
\sim 5$ in the interior (although the radial variation of $\beta$ is
quite different).}
\end{figure}

\end{document}